# Common revenue allocation in DMUs with two stages based on DEA cross-efficiency and cooperative game

Xinyu Wang, Qianwei Zhang[*], Yilun Lu, Yingdi Zhao

School of Mathematics, Renmin University of China, Beijing 100872, PR China

Emails: wangxinyu3@ruc.edu.cn, zqw2002@ruc.edu.cn, lyl_2021103698@ruc.edu.cn, zhaoyingdi@ruc.edu.cn

**Abstract** In this paper, we examine two-stage production organizations as decision making units (DMUs) that can collaborate to form alliances. We present a novel approach to transform a grand coalition of $n$ DMUs with two-stage structure into $2n$ single-stage sub-DMUs by extending the vectors of the initial input, intermediate product, and final output, thus creating a $2n \times 2n$ DEA cross-efficiency (CREE) matrix. By combing cooperative game theory with CREE and utilizing three cooperative game solution concepts, namely, the nucleolus, the least core and the Shapley value, a characteristic function is developed to account for two types of allocation, i.e., direct allocation and secondary allocation. Moreover, the super-additivity and the core non-emptiness properties are explored. It is found that the sum of the revenue allocated to all DMUs will remain constant at each stage regardless of the allocation manner and the cooperative solution concept selected. To illustrate the efficiency and practicality of the proposed approach, both a numerical example and an empirical application is provided.

## 1. Introduction

## 2. 99The mathematical model

In this section, we propose two allocation manners for distributing the revenue in the two-stage series structure: the direct allocation and the secondary allocation modes.

### 2.1 The transformation of DMUs

We consider the allocation problem in a production system consisting of $n$ homogeneous DMUs with the two-stage network structure, as depicted in Figure 1.

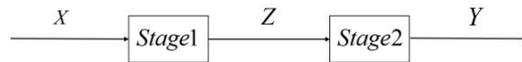

**Figure 1** Two-stage network structure

Each evaluated $DMU_j$ ($j \in \{1, 2, \ldots, n\}$) contains two sub-DMUs, referred to as stage 1 and stage 2, respectively. Here, stage 1 consumes $s$ kinds of initial inputs and produces $q$ kinds of intermediate products. These intermediate products are also the inputs for stage 2, which are used to produce $t$ kinds of final outputs. The input, the intermediate product and the final output vectors of $DMU_j$ are denoted as $\tilde{X}_j = (\tilde{x}_{1j}, \tilde{x}_{2j}, \ldots, \tilde{x}_{sj})$, $\tilde{Z}_j = (\tilde{z}_{1j}, \tilde{z}_{2j}, \ldots, \tilde{z}_{qj})$ and $\tilde{Y}_j = (\tilde{y}_{1j},

$\tilde{y}_{2j}, \ldots, \tilde{y}_{tj}$) ($j \in \{1, 2, \ldots, n\}$), respectively.

In the subsequent common revenue allocation process, to account for the size of each DMU, the normalized input-output measures are utilized. To achieve this, the input, the intermediate product and the final output data are converted to the measures that can be utilized to identify the size parameters (Li, Zhu et al. 2019):

$$x_{ij} = \frac{\tilde{x}_{ij}}{\sum_{j=1}^{n} \tilde{x}_{ij}}, \quad z_{rj} = \frac{\tilde{z}_{rj}}{\sum_{j=1}^{n} \tilde{z}_{rj}}, \quad y_{kj} = \frac{\tilde{y}_{kj}}{\sum_{j=1}^{n} \tilde{y}_{kj}}, \quad \forall i, r, k, j.$$

By extending the vectors of the initial input, the intermediate product, and the final output, we propose an innovative transformation method to convert $n$ DMUs with two-stage series structure into $2n$ sub-DMUs. Specifically, We extend the dimensionality of all the inputs, the intermediate products, and the final outputs by adding the intermediate product vectors to both the input vectors of stage 1 and the output vectors of stage 2. In this way, each sub-DMU has a single-stage structure and uses $(X, Z)_{s+q}$ inputs to produce $(Z, Y)_{q+t}$ outputs, which transforms the original production procedure $X \to Z \to Y$ into two processes: $(X, 0)_{s+q} \to (Z, 0)_{q+t}$ for the sub-DMU in stage 1 and $(0, Z)_{s+q} \to (0, Y)_{q+t}$ for the sub-DMU in stage 2, respectively. Note that these zeros represent the zero vectors in the corresponding dimension.

In fact, by calculating the cross-efficiencies among these $2n$ DMUs after the redefinition, it is evident that when the evaluating DMUs and the targeted DMUs belong to different stages, the CREE value between them is generally 0. It is easy to deduce that these meaningless evaluations will be eliminated by our calculation algorithm, leaving only the expected mutual results between the same stages. Additionally, we can calculate the cross-efficiencies among the $n$ DMUs of stage 1 and the cross-efficiencies of stage 2. Therefore, it can be concluded that the CREE can be calculated from two different perspectives, leading to two corresponding revenue allocation schemes later.

Next, we will present the cross-efficiency matrix of the $2n$ DMUs, and discuss its uniqueness problem.

## 2.2 Cross-efficiency and uniqueness problem

In conventional DEA models, the efficiency of each DMU is determined by its self-assessment (Wu, Chu et al. 2016), which involves selecting a set of optimal input and output multipliers (which are typically not unique) to evaluate each DMU. To counter this issue, Sexton et al. (1986) proposed the CREE model, which is a fair and reasonable mechanism for assessing the relative effectiveness of DMUs. The core idea behind the CREE model is the peer evaluation. The CREE of the evaluated DMU is obtained from the optimal weights of the other DMUs.

We consider the $2n$ sub-$DMU_l$ ($l = 1, 2, \cdots, 2n$) mentioned in Subsection 2.1. The input and output vectors of sub-$DMU_l$ are denoted as $\bar{X}_l = (\bar{x}_{1l}, \bar{x}_{2l}, \cdots, \bar{x}_{(s+q)l})$ and $\bar{Y}_l = (\bar{y}_{1l}, \bar{y}_{2l}, \cdots, \bar{y}_{(q+t)l})$, respectively. The CCR efficiency value of the evaluated sub-$DMU_d$ ($d = 1, 2, \cdots, 2n$) can be calculated as follows,

$$max \sum_{r=1}^{q+t} \mu_r \bar{y}_{rd} = \theta_d$$

$$s.t. \quad \sum_{r=1}^{q+t} \mu_r \bar{y}_{rl} - \sum_{k=1}^{s+q} \omega_k \bar{x}_{kl} \leq 0, l = 1, 2, \cdots, 2n, \quad (1)$$

$$\sum_{k=1}^{s+q} \omega_k \bar{x}_{kd} = 1,$$

$$\mu_r \geq 0, \omega_k \geq 0, r = 1, 2, \cdots, q+t; k = 1, 2, \cdots, s+q.$$

For each evaluated sub-$DMU_d$ $(d = 1,2,\cdots,2n)$, we can obtain a set of optimal weights denoted as $\omega^{d^*} = (\omega_1^{d^*}, \omega_2^{d^*}, \cdots, \omega_{s+q}^{d^*})$ and $\mu^{d^*} = (\mu_1^{d^*}, \mu_2^{d^*}, \cdots, \mu_{q+t}^{d^*})$, respectively, the corresponding optimal value of which is $\theta_d^*$. Obviously, these optimal weights are most preferred by $sub-DMU_d$. By using the optimal weights of sub-$DMU_d$, we can obtain the CREE value of any $sub-DMU_l$ ($l = 1,2,\cdots,2n$), as expressed by the following formula (2):

$$E_{d,l} = \frac{\mu^{d^*}\overline{Y}_l^T}{\omega^{d^*}\overline{X}_l^T} = \frac{\sum_{r=1}^{q+t}\mu_r^{d^*}\bar{y}_{rl}}{\sum_{k=1}^{s+q}\omega_k^{d^*}\bar{x}_{kl}}, \quad l = 1,2,\cdots,2n. \quad (2)$$

The above formula represents the evaluation efficiency value of $sub-DMU_d$ for $sub-DMU_l$, with $E_{d,l}$ being termed the $d$-cross-efficiency of $sub-DMU_l$.

However, the optimal solution of model (1) is not unique, and thus the CREE value obtained by formula (2) is not unique either. To present our method and acquire the unique revenue allocation results, we use the new aggressive CREE model proposed by Wu et al. (2016). On the basis of the traditional CCR model, Wu et al. maximized the self-evaluation efficiency value (i.e., the value of $E_{d,d} = \mu^{d^*}\overline{Y}_d^T = \theta_d^*$), and then minimized the CREE value of $E_{d,l} = \frac{\mu^{d^*}\overline{Y}_l^T}{\omega^{d^*}\overline{X}_l^T}$. Thus, this model is based on the premise of maximal efficiency $\theta_d^*$, followed by minimizing the CREE value of $E_{d,l}$. The specific model to calculate the unique CREE of $E_{d,l}$ is presented below,

$$\min E_{d,l} = \mu^d\overline{Y}_l^T$$
$$s.t. \quad \mu^d\overline{Y}_l^T - \omega^d\overline{X}_l^T \leq 0, l = 1,2,\cdots,2n, \quad (3)$$
$$\omega^d\overline{X}_l^T = 1,$$
$$\theta_d^*\omega^d\overline{X}_l^T - \mu^d\overline{Y}_l^T = 0,$$
$$\omega^d \geq 0, \mu^d \geq 0.$$

By solving the linear programming problem above, we can obtain the unique value $E_{d,l} = \mu^{d^*}\overline{Y}_l^T$ according to the optimal weights $\mu^{d^*}$ and $\omega^{d^*}$. As the $sub-DMU_l$ ($l = 1,2,\cdots,2n$) has $2n$ $d$-cross-efficiency values, upon calculating the model (3) $4n^2$ times, the unique CREE matrix in $2n \times 2n$ dimensions can be obtained, as depicted in Table 1.

**Table 1** A generalized cross-efficiency matrix (CEM)

| Evaluator $DMU_j$ | Targeted DMU | | | | |
|---|---|---|---|---|---|
| | 1 | 2 | 3 | ... | n |
| 1 | $E_{11}$ | $E_{12}$ | $E_{13}$ | ... | $E_{1n}$ |
| 2 | $E_{21}$ | $E_{22}$ | $E_{23}$ | ... | $E_{2n}$ |
| 3 | $E_{31}$ | $E_{32}$ | $E_{33}$ | ... | $E_{3n}$ |
| ... | ... | ... | ... | ... | ... |
| n | $E_{n1}$ | $E_{n2}$ | $E_{n3}$ | ... | $E_{nn}$ |
| ... | ... | ... | ... | ... | ... |
| 2n | $E_{2n1}$ | $E_{2n2}$ | $E_{2n3}$ | ... | $E_{2n2n}$ |

Theorem 1 presents us with the following conclusion concerning the CREE values of these $2n$ homogeneous subjects.

**Theorem 1.** The cross-efficiency value of mutual evaluation in different stages is equal to 0.

**Proof.** According to our design, both stages of all DMUs can be transformed into $2n$ sub-DMUs, having the following structure:

$$(X,Z)_{s+q} \to (Z,Y)_{q+t}.$$

By taking $DMU_A$ and $DMU_B$ arbitrarily from the original set of $n$ DMUs, we consider the CREE values between $DMU_{A.1}$ (stage 1 of $DMU_A$), $DMU_{B.1}$ (stage 1 of $DMU_B$), $DMU_{A.2}$ (stage 2 of $DMU_A$), and $DMU_{B.2}$ (stage 2 of $DMU_B$). According to the symmetry, we only need to prove that the mutual evaluation (CREE) between $DMU_{A.1}$ and $DMU_{B.2}$ is 0, and the same is also true for the evaluation between $DMU_{A.2}$ and $DMU_{B.1}$.

For stage 1, the production process is $(X,0)_{s+q} \to (Z,0)_{q+t}$, and for stage 2, it is $(0,Z)_{s+q} \to (0,Y)_{q+t}$. According to the model (1), we obtain the optimal efficiency and the weights of $DMU_{A.1}$, denoted as $\theta_{A.1}^*$, $\omega^{A.1^*}$ and $\mu^{A.1^*}$, respectively. $\omega^{A.1^*}$ and $\mu^{A.1^*}$ can further be specified as

$$\omega^{A.1^*} = (\omega_1^{A.1^*}, \omega_2^{A.1^*}, \cdots, \omega_{s+q}^{A.1^*}) = (\omega_1^{A.1^*}, \omega_2^{A.1^*}, \cdots, \omega_s^{A.1^*}, \widetilde{\omega}_1^{A.1^*}, \widetilde{\omega}_2^{A.1^*}, \cdots, \widetilde{\omega}_q^{A.1^*}) = (\overrightarrow{\omega_s}^{A.1^*}, \overrightarrow{\omega_q}^{A.1^*})$$

and

$$\mu^{A.1^*} = (\mu_1^{A.1^*}, \mu_2^{A.1^*}, \cdots, \mu_{q+t}^{A.1^*}) = (\mu_1^{A.1^*}, \mu_2^{A.1^*}, \cdots, \mu_q^{A.1^*}, \widetilde{\mu}_1^{A.1^*}, \widetilde{\mu}_2^{A.1^*}, \cdots, \widetilde{\mu}_t^{A.1^*}) = (\overrightarrow{\mu_q}^{A.1^*}, \overrightarrow{\mu_t}^{A.1^*}),$$

respectively, with $\overrightarrow{\omega_s}^{A.1^*} = (\omega_1^{A.1^*}, \omega_2^{A.1^*}, \cdots, \omega_s^{A.1^*})$, $\overrightarrow{\omega_q}^{A.1^*} = (\widetilde{\omega}_1^{A.1^*}, \widetilde{\omega}_2^{A.1^*}, \cdots, \widetilde{\omega}_q^{A.1^*})$, $\overrightarrow{\mu_q}^{A.1^*} = (\mu_1^{A.1^*}, \mu_2^{A.1^*}, \cdots, \mu_q^{A.1^*})$ and $\overrightarrow{\mu_t}^{A.1^*} = (\widetilde{\mu}_1^{A.1^*}, \widetilde{\mu}_2^{A.1^*}, \cdots, \widetilde{\mu}_t^{A.1^*})$.

Then, we can calculate the CREE using formula (2):

$$E_{A.1,B.2} = \frac{\mu^{A.1^*} Y_{B.2}^T}{\omega^{A.1^*} X_{B.2}^T} = \frac{\overrightarrow{\mu_q}^{A.1^*} \cdot \overrightarrow{0_q} + \overrightarrow{\mu_t}^{A.1^*} \cdot Y_t^T}{\overrightarrow{\omega_s}^{A.1^*} \cdot \overrightarrow{0_s} + \overrightarrow{\omega_q}^{A.1^*} \cdot Z_q^T} = \frac{\overrightarrow{\mu_t}^{A.1^*} \cdot Y_t^T}{\overrightarrow{\omega_q}^{A.1^*} \cdot Z_q^T}.$$

Furthermore, we adopt the "aggressive CREE model" (Wu et al. 2016) to solve the uniqueness problem of CREE, and choose the weights to minimize the CREE value $E_{A.1,B.2}$. To this end, there exists a vector $\overrightarrow{\mu_t}^{A.1^*} = (0, \cdots, 0)$ such that the CREE can reach the minimum value $E_{A.1,B.2} = 0$.

Therefore, we have proven that the CREE value of mutual evaluation in different stages is equal to 0.

Assuming that these $2n$ sub-DMUs can cooperate with each other to form coalitions, the characteristic function of the cooperative game can be establish based on CREE, which is discussed in the following Subsection 2.3.

## 2.3 Characteristic function

In this subsection, according to the CREE matrix, we design a characteristic function to reflect a fair distribution of revenues with the alliance.

Suppose the coalition $S$ is a subset of the grand set of participants $N = \{1,2,\cdots,2n\}$. For any participant $i$ belonging to the coalition $S$, we define the CREE for the participant $i$ of the alliance $S$ by the following definition.

**Definition 1.** Let the coalition $S$ ($|S| \geq 2$) be a subset of $N = \{1,2,\cdots,2n\}$, and let $|S|$ represent the number of participants in the coalition $S$. For any participant $i$ belonging to the coalition $S$, the CREE of $i$ is defined as follows:

$$e_{S,i}^{cross} = \max_{d \in S}\{E_{d,i}, d \neq i\}. \qquad (4)$$

The CREE of the participant $i$ in the coalition $S$ is defined as the maximum value of all the cross-efficiencies $E_{d,i}$, where $d \in S$ and $d \neq i$. When there is only one participant $i$ in the coalition $S$, i.e., the participant $i$ does not cooperate with any other DMUs, the CREE of the participant $i$ is defined as the minimum value of CREE $E_{d,i}$, where $d \in N$ and $d \neq i$, which can be expressed by the following formula (5),

$$e_i^{cross} = \min_{d \in N}\{E_{d,i}, d \neq i\}. \quad (5)$$

According to reference (Osborne et al. 1994), a coalition game can be expressed by the following definition.

**Definition 2.** Let $S$ be a non-empty subset of a set of players $N = \{1,2,\cdots,2n\}$. A coalition game with transferable utility (TU-game) $\langle N, v \rangle$ consists of two components: $N$, a finite set of players (a grand coalition), and $v(S)$, a characteristic function which assigns a real number to each non-empty subset $S$ of $N$, reflecting its worth.

We provide a definition of the characteristic function of a coalition game $\langle N, v \rangle$ based on CREE, as outlined in the following Definition 3, to address the issue of revenue allocation.

**Definition 3.** The characteristic function $v(S)$ for the coalition $S \subseteq N$ is defined by formula (6),

$$v(S) = \frac{f(S)R}{f(N)}, \quad (6)$$

where $f(S) = \sum_{i \in S} e_{S,i}^{cross}$ represents the sum of the cross-efficiency values of all the participants in the coalition $S$, and $R$ is the revenue of the grand coalition $N$ which needs to be allocated.

If there is only one participant $i$ in the coalition $S$, then we have $v(i) = \frac{f(i)R}{f(N)}$, where $f(i) = e_i^{cross}$.

In formula (6), the $\frac{f(S)}{f(N)}$ represents the proportion of the sum of CREE values in the coalition $S$ to that of the grand alliance $N$, thereby reflecting the value of the coalition $S$ relative to the grand coalition $N$ through this proportional relationship.

## 3. Properties and the revenue allocation procedure

In this section, we discuss the properties of the characteristic function of the coalition game and present two revenue allocation manners, which will be illustrated in the procedure figure at the end of this section.

### 3.1 Properties of the characteristic function

In this subsection, we discuss two properties of the characteristic function, namely super-additivity and core non-emptiness.

#### 3.1.1 The property of super-additivity

In the coalition game, there is a basic assumption that the characteristic function $v(S)$ satisfies the property of super-additivity, which is the fundamental condition for the formation of an alliance. To that end, we first define the super-additivity, and then investigate whether the characteristic function given by us fulfills this property.

**Definition 4.** The characteristic function $v(S)$ satisfies the property of super-additivity, meaning that for any coalitions $S$ and $T$ with $S \cap T = \emptyset$, it holds that

$$v(S \cup T) \geq v(S) + v(T).$$

The super-additivity of the characteristic function implies that the revenue of a larger coalition is not less than the combined revenue of its two sub-coalitions. Thus, the bigger the alliance, the greater the revenue it will generate, which is the fundamental rationale behind the formation of alliances.

The characteristic function $v(S)$ proposed by us satisfies the property of super-additivity and can be expressed by the following Theorem 2.

**Theorem 2.** The characteristic function $v(S)$ of formula (6) is super-additive, i.e., for any $S_1 \subseteq N$, $S_2 \subseteq N$ with $S_1 \cap S_2 = \emptyset$, it holds that

$$v(S_1 \cup S_2) \geq v(S_1) + v(S_2).$$

**Proof.** According to the formula (6), we have

$$v(S) = \frac{f(S)R}{f(N)}.$$

It is obvious that $\frac{R}{f(N)}$ is a constant. Therefore, in order to demonstrate the super-additivity of the characteristic function $v(S)$, we only need to show that $f(S)$ is super-additive. According to Definitions 1 and 3, for any $S_1 \subseteq N$, $S_2 \subseteq N$ with $S_1 \cap S_2 = \emptyset$, we have

$$f(S_1) = \sum_{i \in S_1} e_{S_1,i}^{cross} = \sum_{i=1}^{s_1} e_{S_1,i}^{cross} = \sum_{i=1}^{s_1} \max_{d \in S_1} \{E_{d,i}, d \neq i\}$$

and

$$f(S_2) = \sum_{i \in S_2} e_{S_2,i}^{cross} = \sum_{i=1}^{s_2} e_{S_2,i}^{cross} = \sum_{i=1}^{s_2} \max_{d \in S_2} \{E_{d,i}, d \neq i\},$$

where the numbers of players in coalitions $S_1$ and $S_2$ are denoted as $s_1$ and $s_2$, respectively. Then, we have

$$f(S_1 \cup S_2) = \sum_{i \in S_1 \cup S_2} e_{S_1 \cup S_2,i}^{cross} = \sum_{i=1}^{s_1+s_2} e_{S_1 \cup S_2,i}^{cross} = \sum_{i=1}^{s_1+s_2} \max_{d \in S_1 \cup S_2} \{E_{d,i}, d \neq i\}.$$

Since for any $i \in S_1 \cup S_2$, there are

$$\max_{d \in S_1} \{E_{d,i}, d \neq i\} \leq \max_{d \in S_1 \cup S_2} \{E_{d,i}, d \neq i\}$$

and

$$\max_{d \in S_2} \{E_{d,i}, d \neq i\} \leq \max_{d \in S_1 \cup S_2} \{E_{d,i}, d \neq i\},$$

we can see that the following equalities and inequalities are valid,

$$f(S_1 \cup S_2) = \sum_{i=1}^{s_1+s_2} \max_{d \in S_1 \cup S_2} \{E_{d,i}, d \neq i\}$$

$$= \sum_{i=1}^{s_1} \max_{d \in S_1 \cup S_2} \{E_{d,i}, d \neq i\} + \sum_{i=1}^{s_2} \max_{d \in S_1 \cup S_2} \{E_{d,i}, d \neq i\}$$

$$\geq \sum_{i=1}^{s_1} \max_{d \in S_1} \{E_{d,i}, d \neq i\} + \sum_{i=1}^{s_2} \max_{d \in S_2} \{E_{d,i}, d \neq i\}$$

$$= f(S_1) + f(S_2).$$

Therefore, the super-additivity of $f(S)$ has been proved. Consequently, the characteristic function $v(S)$ satisfies

$$v(S_1 \cup S_2) \geq v(S_1) + v(S_2).$$

### 3.1.2 The property of core non-emptiness

Let $\langle N, v \rangle$ be a cooperative game. The vector $\hat{x} = (x_1, x_2, \cdots, x_{2n})$ is an imputation of $\langle N, v \rangle$ if it satisfies

$$\sum_{i=1}^{2n} x_i = v(N) \tag{7}$$

and

$$x_i \geq v(i), \quad i = 1, 2, \cdots, 2n. \tag{8}$$

The formulas (7) and (8) indicate the collective and the individual rationalities, respectively. We introduce the conception of the core by Definition 5.

**Definition 5.** The core is the set of all imputations, so it is a set-valued solution of the cooperative game, i.e.,

$$core\langle N, v \rangle := \{\hat{x} = (x_1, x_2, \cdots, x_{2n}) | \sum_{i \in N} x_i = v(N)\} \tag{9}$$

and

$$\sum_{i \in S} x_i \geq v(S) \text{ for all } S \subset N. \tag{10}$$

It is not difficult to see that there is no other solution that can dominate the imputations in the core.

**Theorem 3.** For the characteristic function $v(S)$ of formula (6), there exists an imputation of the cooperative game $\langle N, v \rangle$ that satisfies both the individual and the collective rationalities. In other words, there is at least one imputation $\hat{x} = (x_1, x_2, \cdots, x_{2n})$ such that

$$\sum_{i \in N} x_i = v(N), \quad x_i \geq v(i), \ i = 1, 2, \cdots, 2n, \tag{11}$$

i.e., $core\langle N, v \rangle \neq \emptyset$.

**Proof.** Suppose that the revenue to be allocated is $R$, i.e., $\sum_{i \in N} x_i = R$. Since

$$v(S) = \frac{f(S)R}{f(N)}, \quad v(i) = \frac{f(i)R}{f(N)}, \quad i = 1, 2, \cdots, n,$$

we have

$$v(N) = \frac{f(N)R}{f(N)} = R = \sum_{i \in N} x_i.$$

This equation reveals that the distribution $\hat{x} = (x_1, x_2, \cdots, x_n)$ is collectively rational. Moreover, due to the super-additivity of $f$, we obtain

$$f(N) \geq \sum_{i \in N} f(i) = \sum_{i=1}^{2n} f(i).$$

Therefore,

$$\frac{R \sum_{i=1}^{2n} f(i)}{f(N)} \leq \frac{Rf(N)}{f(N)} = R = \sum_{i \in N} x_i.$$

Because of $v(i) = \frac{f(i)R}{f(N)}$, it holds

$$\sum_{i \in N} x_i \geq \frac{R}{f(N)} \sum_{i=1}^{2n} f(i) = \sum_{i=1}^{2n} \frac{Rf(i)}{f(N)} = \sum_{i=1}^{2n} v(i).$$

Due to the property of non-negativity of $x_i$ and $v(i)$, $i = 1, 2, \cdots, 2n$, there must exist an

$n$-dimension vector such that $x_i \geq v(i)$, which implies that the allocation result $\hat{x} = (x_1, x_2, \cdots, x_{2n})$ satisfies individual rationality.

According to Definition 5, we can conclude that the core is non-empty.

Theorem 3 reveals that the cooperative game $\langle N, v \rangle$ has a non-empty core, which confirms the existence of the nucleolus. Therefore, taking into account the satisfaction of the participants, we can utilize the nucleolus to allocate the common revenue among the $2n$ sub-DMUs and obtain an equitable and unique allocation scheme to guarantee the stability of the alliance.

**3.2 The procedure of the revenue allocation scheme**

In cooperative games, there are many solutions for distributing the surplus value of the alliance from different perspectives, including the nucleolus, the least core, and the Shapley value. This subsection introduces these concepts and then expounds upon two partitioning approaches to the common revenue and the properties of each method under the three solution concepts of the cooperative game.

**3.2.1 The nucleolus, the least core, and the Shapley value**

Based on the Definition 5, the conception of the nucleolus can be described as follows.

**Definition 6** (Luo, Zhou et al. 2022). The nucleolus is defined as the imputation $\hat{x} = (x_1, x_2, \cdots, x_{2n})$ such that the excess

$$e_S(\hat{x}) = v(S) - \sum_{i \in S} x_i \qquad (12)$$

of any possible coalition $S \subseteq N$ cannot be reduced without increasing any other greater excess. Here $e_S(\hat{x})$ measures the satisfaction of the various players in the grand coalition with the allocation scheme.

Furthermore, Schmeidler (1969) proposed that a non-empty core of a cooperative game must have a unique nucleolus. The existence of the nucleolus of the cooperative game proposed by us can be guaranteed by Theorem 3.

The least core is constructed based on the core and the nucleolus and is an alternative solution based on the satisfaction. Additionally, when the core of the cooperative game is empty, its least core still exists (Kern and Paulusma 2003).

**Definition 7.** The least core based on the characteristic function $v(S)$ can be expressed by the following programming:

$$\max \varepsilon$$
$$s.t. \begin{cases} x(S) \geq v(S) + \varepsilon, & S \notin \{\emptyset, N\}, \\ x(N) = v(N), & otherwise. \end{cases} \qquad (13)$$

It is obvious that the optimal value $\varepsilon^* \geq 0$, if and only if $\text{core}\langle N, v \rangle \neq \emptyset$.

In fact, for cooperative games with a non-empty core, the least core can be determined using the iterative method to obtain the nucleolus.

Proposed from the perspective of probability theory, the Shapley value reflects the extent to which each participant contributes to the alliance, and it has been widely used due to its uniqueness similar to that of the nucleolus and its simple calculation process. According to the cooperative game theory, we know that the Shapley value, based on the characteristic function $v(S)$ is defined by Definition 8 (Shapley 1953).

**Definition 8.** The Shapley value based on the characteristic function $v(S)$ can be calculated by using the following formula:

$$\varphi_i(v) = \sum_{S \subset N, i \notin S} \frac{s!(2n-s-1)!}{2n!} \cdot [v(S \cup \{i\}) - v(S)], i \in N, \quad (14)$$

where for any $i \in N$, $s$ and $2n$ are the numbers of participants in coalitions $S$ and $N$, respectively.

The value of the above formula indicates the contribution of the participant $i$ in a cooperative game. For an $2n$-person game, the Shapley value is expressed by a $2n$-dimensional vector $(\varphi_1(v), \varphi_2(v), \cdots, \varphi_{2n}(v))$, where each component denotes the extent to which a corresponding participant contributed to the grand coalition $N$. For the revenue allocation problem, the Shapely value, combined with the peer evaluation character of CREE, can be used to obtain the revenue allocation scheme expressed as

$$x_i = \frac{\varphi_i(v)R}{\sum_{i=1}^{n} \varphi_i(v)}, \quad i = 1, 2, \cdots, 2n, \quad (15)$$

where $x_i$ is the revenue obtained by the $i$-th sub-DMU and $\hat{x} = (x_1, x_2, \cdots, x_{2n})$ is the vector representing the revenue allocation result among the $2n$ sub-DMUs.

### 3.2.2 Two allocation schemes of the common revenue

Next, we introduce two allocation schemes of the common revenue between the DMUs and the stages within the DMUs.

**Direct allocation mode:** the common revenue is allocated directly among the $2n$ sub-DMUs which are derived from the $n$ DMUs, as discussed in Subsection 2.1.

**Secondary allocation mode:** for the $n$ DMUs with two-stage series structure, denoting the grand alliance formed by stage 1 as $N_1$ and the alliance formed by stage 2 as $N_2$, we allocate the common revenue between $N_1$ and $N_2$ firstly, and then redistribute the allocated revenue among the internal $n$ sub-DMUs of each stage.

These two allocation schemes are presented visually in Figure 2.

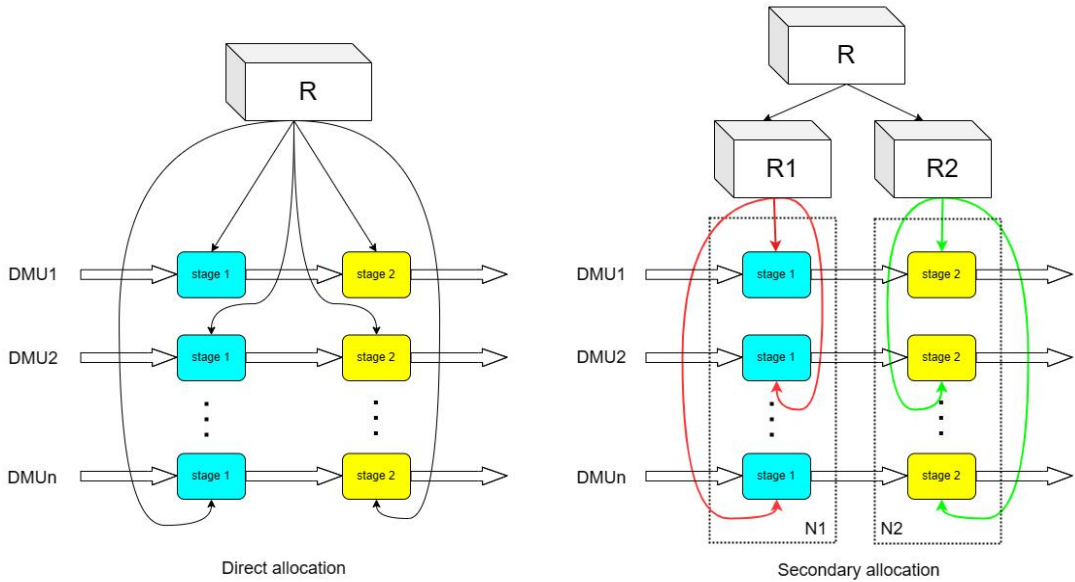

**Figure 2** Calculation flow chart of the proposed model

When the revenue is allocated within the alliances using these two allocation manners and different conceptions of the cooperative game, we can draw the following conclusions.

**Theorem 4.** The common revenue distributed to each sub-DMU remains unchanged according to the nucleolus concept, regardless of the direct or the secondary allocation method is chosen.

**Proof.** Suppose that $N = \{1, 2, \cdots, 2n\}$, and the vector $\hat{x} = (x_1, x_2, ..., x_n, x_{n+1}, ..., x_{2n})$ is the nucleolus for the cooperative game $\langle N, v \rangle$. We denote

$$\hat{x}_1 = (x_1, x_2, ..., x_n), \quad \hat{x}_2 = (x_{n+1}, x_{n+2}, ..., x_{2n})$$

According to Definition 6, we know that for any possible coalition $S \subseteq N$, it is impossible to reduce the excess of $e_S(\hat{x}) = v(S) - \sum_{i \in S} x_i$ without increasing any other greater excess.

Furthermore, we can conclude that the excesses of $e_{S_1}(\hat{x}) = v(S_1) - \sum_{i \in S_1} x_i$ and $e_{S_2}(\hat{x}) = v(S_2) - \sum_{i \in S_2} x_i$ cannot be reduced without increasing any other greater excess for any potential coalitions $S_1 \subseteq N_1$ and $S_2 \subseteq N_2$, where $N_1$ denotes the alliance formed by stage 1 and $N_2$ represents the alliance formed by stage 2, respectively.

For the secondary allocation manner, if $\hat{y} = (y_1, y_2, ..., y_n)$ and $\hat{z} = (z_1, z_2, ..., z_n)$ are the nucleolus for the cooperative games $\langle N_1, v \rangle$ and $\langle N_2, v \rangle$, respectively, then the excess of $e_{S_1}(\hat{y}) = v(S_1) - \sum_{i \in S_1} y_i$ for any coalitions $S_1 \subseteq N_1$ cannot be reduced without increasing any other greater excess. The same is also true for $e_{S_2}(\hat{z}) = v(S_2) - \sum_{i \in S_2} z_i$ and any coalition $S_2 \subseteq N_2$.

Since the nucleolus is unique, it is obvious that $x_i = y_i (i = 1, 2, ..., n)$ and $x_{n+j} = z_j (j = 1, 2, ..., n)$. These equations mean that the common revenue $R$ distributed in the nucleolus remains unchanged for each sub-DMU in both the direct and the secondary allocation manners.

Furthermore, it is evident that $R_1 = \sum_{i \in N_1} x_i = \sum_{i \in N_1} y_i$, $R_2 = \sum_{i \in N_2} x_i = \sum_{i \in N_2} y_i$ and $R_1 + R_2 = R$.

For the two other cooperative game solution concepts, the least core and the Shapley value, the common revenue allocation results remain unchanged in the two allocation manners, as expressed in Theorem 5.

**Theorem 5.** The allocated revenues of the coalitions $N_1$ and $N_2$ remain constant no matter which allocation method or cooperative game solution concepts--- the nucleolus, the least core or the Shapley value---are adopted, i.e., $R_1 = \sum_{i \in N_1} x_i$ and $R_2 = \sum_{i \in N_2} x_i$ are fixed values for a given total revenue $R$.

**Proof.** Since the three cooperative game solution concepts of the nucleolus, the least core, and the Shapley value are all based on the same characteristic function $v(S)$, Theorem 1 shows that the CREE value of mutual evaluation in different stages is equal to 0. Consequently, for the characteristic function $v(S)$ defined by Definition 3, we have

$$v(F \cup S) = v(F) + v(S), \text{ for any } F \subseteq N_1, S \subseteq N_2,$$

and in particular,

$$v(N) = v(N_1) + v(N_2).$$

Specifically, for any $F \subseteq N_1, S \subseteq N_2$, we have

$$v(F \cup S) = \sum_{i \in F \cup S} e_{F \cup S,i}^{cross} = \sum_{i \in F \cup S} \max_{d \in F \cup S} \{E_{d,i}, d \neq i\}$$

$$= \sum_{i \in F} \max \left\{ \max_{d \in F} \{E_{d,i}, d \neq i\}, \max_{d \in S} \{E_{d,i}, d \neq i\} \right\}$$

$$+ \sum_{i \in S} \max \left\{ \max_{d \in F} \{E_{d,i}, d \neq i\}, \max_{d \in S} \{E_{d,i}, d \neq i\} \right\}$$

$$= \sum_{i \in F} \max \left\{ \max_{d \in F} \{E_{d,i}, d \neq i\}, 0 \right\} + \sum_{i \in S} \max \left\{ 0, \max_{d \in S} \{E_{d,i}, d \neq i\} \right\}$$

$$= \sum_{i \in F} \max_{d \in F} \{E_{d,i}, d \neq i\} + \sum_{i \in S} \max_{d \in S} \{E_{d,i}, d \neq i\}$$

$$= v(F) + v(S),$$

and especially,
$$v(N) = v(N_1) + v(N_2).$$

For the nucleolus, Theorem 4 states that the common revenue distributed to each sub-DMU remains the same under the two allocation manners, resulting in $R_1 = x(N_1) = v(N_1)$ and $R_2 = x(N_2) = v(N_2)$.

For the least core, according to Definition 7, we have that the following two formulas

$$R_1 = x(N_1) \geq v(N_1) + \varepsilon \tag{16}$$

and

$$R_2 = x(N_2) \geq v(N_2) + \varepsilon \tag{17}$$

hold.

Theorem 3 dictates that the core of the cooperative game $\langle N, v \rangle$ is not empty, and thus $\varepsilon \geq 0$. Consequently, since

$$v(N_1) + v(N_2) = v(N) = x(N) = R = R_1 + R_2 \geq v(N_1) + v(N_2) + 2\varepsilon,$$

this implies that $\varepsilon = 0$, which leads to $R_1 = v(N_1), R_2 = v(N_2)$.

To simplify, according to Definition 8, we normalize $\frac{R}{\sum_{i=1}^{n} \varphi_i(v)} = 1$, resulting in $R = v(N) = \sum_{i \in N} x_i$, and

$$x_i = \varphi_i(v) = \sum_{S \subset N, i \notin S} \frac{s!(2n-s-1)!}{2n!} [v(S \cup \{i\}) - v(S)], \quad i = 1, 2, \cdots, 2n.$$

In the cooperative game $\langle N, v \rangle$, for any subset $S \subset N$, there are the following relationships,

$$\exists S_1 \subset N_1, \ S_2 \subset N_2, \ s.t. \ S_1 \cup S_2 = S, \ S_1 \cap S_2 = \emptyset.$$

In the direct allocation manner, we denote $s_1$ and $s_2$ as the numbers of participants in the coalitions $S_1$ and $S_2$, respectively. Thus, we have

$$R_1 = \sum_{i \in N_1} \varphi_i(v) = \sum_{i \in N_1} \sum_{S \subset N, i \notin S} \frac{s!(2n-s-1)!}{2n!} [v(S \cup \{i\}) - v(S)]$$

$$= \sum_{i \in N_1} \sum_{S \subset N, i \notin S} \frac{s!(2n-s-1)!}{2n!} [v(S_1 \cup S_2 \cup \{i\}) - v(S_1 \cup S_2)]$$

$$= \sum_{i \in N_1} \sum_{S \subset N, i \notin S} \frac{s!(2n-s-1)!}{2n!} [v(S_1 \cup \{i\}) + v(S_2) - v(S_1) - v(S_2)]$$

$$= \sum_{i \in N_1} \sum_{S \subset N, i \notin S} \frac{s!(2n-s-1)!}{2n!}[v(S_1 \cup \{i\}) - v(S_1))]$$

$$= \sum_{i \in N_1} \sum_{S_1 \subset N_1, i \notin S_1} \sum_{S_2 \subset N_2} \frac{s!(2n-s-1)!}{2n!}[v(S_1 \cup \{i\}) - v(S_1))]$$

$$= \sum_{i \in N_1} \sum_{S_1 \subset N_1, i \notin S_1} \sum_{s_2=0}^{n} \binom{n}{s_2} \frac{(s_1+s_2)!(2n-s_1-s_2-1)!}{2n!}[v(S_1 \cup \{i\}) - v(S_1))]$$

$$= \sum_{i \in N_1} \sum_{S_1 \subset N_1, i \notin S_1} [v(S_1 \cup \{i\}) - v(S_1))] \sum_{s_2=0}^{n} \binom{n}{s_2} \frac{(s_1+s_2)!(2n-s_1-s_2-1)!}{2n!}$$

$$= \sum_{i \in N_1} \sum_{S_1 \subset N_1, i \notin S_1} [v(S_1 \cup \{i\}) - v(S_1))] \cdot \frac{s_1!(n-s_1-1)!}{n!} = v(N_1).$$

For the cooperative game $\langle N_1, v \rangle$, through the secondary allocation method of common revenue, we obtain

$$R_1 = x(N_1) = \sum_{i \in N_1} x_i = \sum_{i \in N_1} \sum_{S_1 \subset N_1, i \notin S_1} \frac{s_1!(n-s_1-1)!}{n!}[v(S_1 \cup \{i\}) - v(S_1)].$$

As a result, the value of $R_1 = \sum_{i \in N_1} x_i$ is fixed in both the direct and the secondary allocation methods, and the same statement applies to $R_2$.

Summing up, regardless of the chosen allocation manner and cooperative game concept (the nucleolus, the least core, or the Shapley value), we can observe that the revenue allocated to coalitions $N_1$ and $N_2$ remains constant.

## 4. A numerical example and an empirical application

In this section, we first illustrate our proposed revenue allocation approach through a numerical example, verifying the properties of the allocation results. Afterwards, we apply these approaches to the actual revenue allocation problems in real applications.

### 4.1 A numerical example

We randomly generate 7 DMUs, each with two inputs, an intermediate product, and two outputs. As Table 2 shows, the specific production data of the 7 DMUs is listed. Assuming the total revenue to be allocated is R = 100, we calculate the allocation result of each sub-DMU by using the two proposed allocation manners.

**Table 2** The raw data for the numerical example.

| DMUs | $X_1$ | $X_2$ | $X_3$ | $Z$ | $Y_1$ | $Y_2$ |
|---|---|---|---|---|---|---|
| 1 | 34 | 18 | 38 | 17 | 10 | 33 |
| 2 | 40 | 12 | 33 | 49 | 32 | 15 |
| 3 | 22 | 49 | 21 | 33 | 28 | 27 |
| 4 | 35 | 45 | 18 | 39 | 50 | 48 |
| 5 | 24 | 28 | 45 | 38 | 23 | 35 |
| 6 | 50 | 41 | 31 | 11 | 34 | 16 |
| 7 | 47 | 45 | 50 | 49 | 24 | 11 |

According to the transformation method, the 7 DMUs with the two-stage series structure can be transformed into 14 sub-DMUs. For each sub-DMU, the product procedure is either $(X, 0) \to (Z, 0)$ or $(0, Z) \to (0, Y)$. After the transformation, we can obtain the new input and output data, which are expressed in Table 2. In this table, $z_1$ and $z_2$ are used to indicate the input in stage 2 and the output in stage 1, respectively.

According to model (3), we can obtain a unique CREE matrix of these 14 sub-DMUs; the corresponding cross-efficiencies are listed in Table 3.

**Table 3** The cross-efficiency matrix of 14 DMUS.

| Evaluator $DMU_j$ | Targeted DMU | | | | | | | | | | | | | |
|---|---|---|---|---|---|---|---|---|---|---|---|---|---|---|
| | 1.1 | 1.2 | 2.1 | 2.2 | 3.1 | 3.2 | 4.1 | 4.2 | 5.1 | 5.2 | 6.1 | 6.2 | 7.1 | 7.2 |
| 1.1 | 0.379 | 0 | 1 | 0 | 0.742 | 0 | 0.683 | 0 | 1 | 0 | 0.153 | 0 | 0.697 | 0 |
| 1.2 | 0 | 1 | 0 | 0.158 | 0 | 0.421 | 0 | 0.634 | 0 | 0.474 | 0 | 0.749 | 0 | 0.116 |
| 2.1 | 0.231 | 0 | 1 | 0 | 0.165 | 0 | 0.212 | 0 | 0.332 | 0 | 0.066 | 0 | 0.267 | 0 |
| 2.2 | 0 | 0.19 | 0 | 0.211 | 0 | 0.275 | 0 | 0.415 | 0 | 0.196 | 0 | 1 | 0 | 0.158 |
| 3.1 | 0.304 | 0 | 0.823 | 0 | 1 | 0 | 0.763 | 0 | 0.686 | 0 | 0.15 | 0 | 0.653 | 0 |
| 3.2 | 0 | 1 | 0 | 0.211 | 0 | 0.478 | 0 | 0.72 | 0 | 0.505 | 0 | 1 | 0 | 0.156 |
| 4.1 | 0.206 | 0 | 0.685 | 0 | 0.725 | 0 | 1 | 0 | 0.39 | 0 | 0.164 | 0 | 0.452 | 0 |
| 4.2 | 0 | 1 | 0 | 0.211 | 0 | 0.478 | 0 | 0.72 | 0 | 0.505 | 0 | 1 | 0 | 0.156 |
| 5.1 | 0.316 | 0 | 0.774 | 0 | 0.742 | 0 | 0.683 | 0 | 1 | 0 | 0.139 | 0 | 0.658 | 0 |
| 5.2 | 0 | 1 | 0 | 0.211 | 0 | 0.478 | 0 | 0.72 | 0 | 0.505 | 0 | 1 | 0 | 0.156 |
| 6.1 | 0.294 | 0 | 1 | 0 | 0.744 | 0 | 1 | 0 | 0.539 | 0 | 0.198 | 0 | 0.592 | 0 |
| 6.2 | 0 | 0.19 | 0 | 0.211 | 0 | 0.275 | 0 | 0.415 | 0 | 0.196 | 0 | 1 | 0 | 0.156 |
| 7.1 | 0.378 | 0 | 1 | 0 | 1 | 0 | 0.873 | 0 | 1 | 0 | 0.177 | 0 | 0.764 | 0 |
| 7.2 | 0 | 0.19 | 0 | 0.211 | 0 | 0.275 | 0 | 0.415 | 0 | 0.196 | 0 | 1 | 0 | 0.158 |

In the CREE matrix, it is evident that the cross-efficiencies between different stages are all zero. For example, the cross-efficiencies between stage 1 of DMU1 and stage 2 of DMU2 satisfy $E_{1.1,\ 2.2} = E_{2.2,\ 1.1} = 0$, and those between stage 1 of DMU3 and stage 2 of DMU4 are $E_{3.1,\ 4.2} = E_{4.2,\ 3.1} = 0$. This result is in accordance with the conclusion presented in Theorem 1.

Based on the established characteristic function $v(S) = \frac{f(S)R}{f(N)}$, we distribute the revenue directly among the $2n$ sub-DMUs, and obtain the allocation results according to the different cooperative game conceptions, i.e., the nucleolus, the least core and the Shapley value, respectively. These allocation results are shown in Table 4.

**Table 4** The allocation result of 14 DMUs.

| DMUs | Shapley value | Rank | Least core | Rank | Nucleolus | Rank |
|---|---|---|---|---|---|---|
| 1.1 | 5.43 | 11 | 3.39 | 11 | 4.12 | 11 |
| 1.2 | 9.33 | 4 | 10.72 | 1 | 10.72 | 1 |
| 2.1 | 9.26 | 6 | 10.72 | 1 | 10.68 | 3 |
| 2.2 | 2.54 | 13 | 2.26 | 13 | 2.28 | 13 |
| 3.1 | 9.32 | 5 | 10.72 | 1 | 9.33 | 6 |
| 3.2 | 6.08 | 10 | 5.13 | 10 | 5.13 | 10 |
| 4.1 | 9.05 | 7 | 9.36 | 6 | 10.02 | 5 |
| 4.2 | 8.00 | 8 | 7.72 | 8 | 7.72 | 8 |
| 5.1 | 9.40 | 3 | 10.72 | 1 | 10.38 | 4 |
| 5.2 | 6.18 | 9 | 5.42 | 9 | 5.42 | 9 |
| 6.1 | 3.83 | 12 | 3.26 | 12 | 2.56 | 12 |
| 6.2 | 9.51 | 2 | 10.72 | 1 | 10.72 | 1 |
| 7.1 | 10.03 | 1 | 8.13 | 7 | 9.24 | 7 |
| 7.2 | 2.05 | 14 | 1.71 | 14 | 1.69 | 14 |

The columns 3, 5 and 7 in Table 4 present the rankings of the revenue allocation of the 14 sub-DMUs based on the Shapley value, the least core and the nucleolus, respectively. It is evident that the assignments based on the least core and the nucleolus are similar. These two assignments are considered in terms of the degree of satisfaction of participants. However, from the perspective of participants' contribution to the alliance, the allocation results based on the Shapley value differ from other two types of allocations, particularly for $DMU_{7.1}$. Comparing the input-output data of $DMU_{2.1}$ and $DMU_{7.1}$, it is clear that both DMUs have the same output, $z_2$ (output) = 49. However, $DMU_{2.1}$ consumes less inputs in each category compared with $DMU_{7.1}$, which leads us to the belief that it performs better. This is also confirmed by the average cross-efficiencies of $DMU_{2.1}$ and $DMU_{7.1}$ in Table 3. However, the Shapley value assignment result for $DMU_{7.1}$ (10.03) is larger than that of $DMU_{2.1}$ (9.26), which goes against our intuitive expectations. The cause of this phenomenon is that the Shapley value only considers the contribution of the participants for the alliance, and not the satisfaction of other members. Additionally, it is greatly affected by the changes in the combinations of other members in the alliance.

For intuition, the three assignment results in Table 4 are presented in the form of line plots in Figure 3. It can be observed that the allocation results obtained by using the three solution concepts of the cooperative game are quite similar. Notably, the allocation results based on the nucleolus have superior properties, as it not only considers the satisfaction of all participants and has existence and uniqueness, but also the allocation results of it are very similar to the other two allocation results based on the least core and the Shapley value. This phenomenon indicates that the allocation results based on nucleolus are more acceptable by the 7 DMUs and their sub-stages.

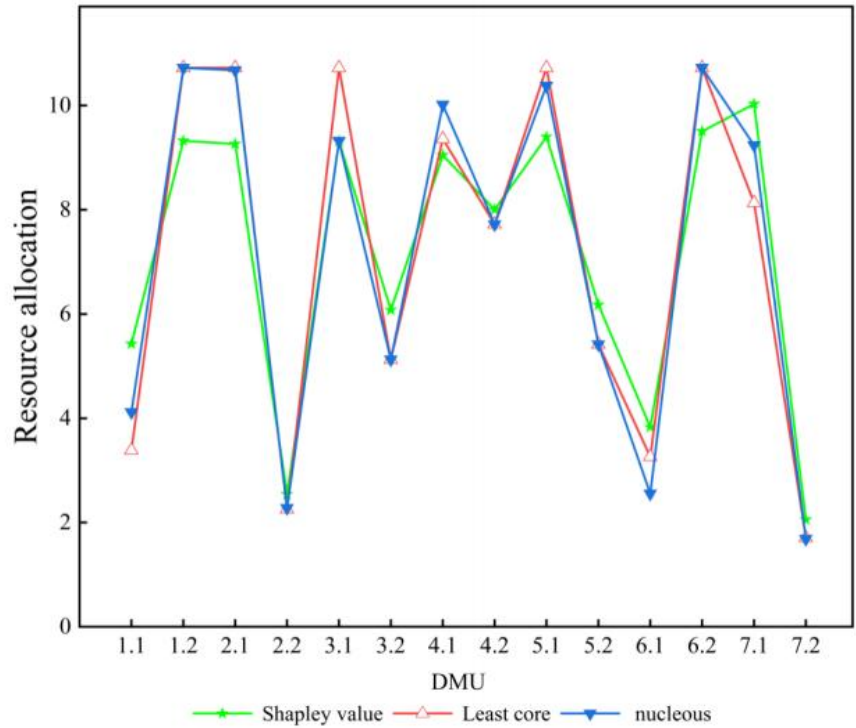

**Figure 3** Resource allocation scheme

Next, we discuss the second distribution mode, namely the secondary distribution pattern. Revenue is first allocated between the first and the second stages of these 7 DMUs, and then redistributed among the internal 7 sub-DMUs of each stage. To this end, we calculate the cross-efficiencies of stage 1 and stage 2, respectively. The resulting data is presented in Tables 5 and 6.

**Table 5** The cross-efficiency matrix of 7 DMUs in stage 1.

| Evaluator $DMU_j$ | Targeted DMU | | | | | | |
| --- | --- | --- | --- | --- | --- | --- | --- |
| | 1.1 | 2.1 | 3.1 | 4.1 | 5.1 | 6.1 | 7.1 |
| 1.1 | 0.38 | 1 | 0.74 | 0.68 | 1 | 0.15 | 0.7 |
| 2.1 | 0.23 | 1 | 0.17 | 0.21 | 0.33 | 0.07 | 0.27 |
| 3.1 | 0.3 | 0.82 | 1 | 0.76 | 0.69 | 0.15 | 0.65 |
| 4.1 | 0.21 | 0.69 | 0.73 | 1 | 0.39 | 0.16 | 0.45 |
| 5.1 | 0.32 | 0.77 | 0.74 | 0.68 | 1 | 0.14 | 0.66 |
| 6.1 | 0.29 | 1 | 0.74 | 1 | 0.54 | 0.2 | 0.59 |
| 7.1 | 0.38 | 1 | 1 | 0.87 | 1 | 0.18 | 0.76 |

**Table 6** The cross-efficiency matrix of 7 DMUs in stage 2.

| Evaluator $DMU_j$ | Targeted DMU | | | | | | |
|---|---|---|---|---|---|---|---|
| | 1.2 | 2.2 | 3.2 | 4.2 | 5.2 | 6.2 | 7.2 |
| 1.2 | 1 | 0.16 | 0.42 | 0.63 | 0.47 | 0.75 | 0.12 |
| 2.2 | 0.19 | 0.21 | 0.28 | 0.42 | 0.2 | 1 | 0.16 |
| 3.2 | 1 | 0.21 | 0.48 | 0.72 | 0.51 | 1 | 0.16 |
| 4.2 | 1 | 0.21 | 0.48 | 0.72 | 0.51 | 1 | 0.16 |
| 5.2 | 1 | 0.21 | 0.48 | 0.72 | 0.51 | 1 | 0.16 |
| 6.2 | 0.19 | 0.21 | 0.28 | 0.42 | 0.2 | 1 | 0.16 |
| 7.2 | 0.19 | 0.21 | 0.28 | 0.42 | 0.2 | 1 | 0.16 |

For simplicity, in order to be consistent with the previous notation, we denote the grand alliances of stages 1 and 2 as $N_1$ and $N_2$, respectively. According to the characteristic function designed by formula (6) and Theorem 1, we can calculate the revenue allocated to $N_1$ and $N_2$ by the equations $R_1 = v(N_1) = \frac{f(N_1)R}{f(N)} = \frac{\Sigma_{i \in N_1} e^{cross}_{N_1,i}}{\Sigma_{i \in N_1} e^{cross}_{N_1,i} + \Sigma_{i \in N_2} e^{cross}_{N_2,i}} \times R$ and $R_2 = v(N_2) = \frac{f(N_2)R}{f(N)} = \frac{\Sigma_{i \in N_2} e^{cross}_{N_2,i}}{\Sigma_{i \in N_1} e^{cross}_{N_1,i} + \Sigma_{i \in N_2} e^{cross}_{N_2,i}} \times R$, respectively. Since R = 100, as shown in Tables 6 and 7, we have $R_1 = 56.32$ and $R_2 = 43.68$.

Furthermore, according to the Shapley value, the least core, and the nucleolus, we redistribute the revenues $R_1$ and $R_2$ among the sub-DMUs of stages 1 and 2, respectively. The final allocation results are presented in the second to fourth columns of Table 7.

Table 7 The second allocation result of 14 DMUs.

| DMUs | Shapley value | Rank | Least core | Rank | Nucleolus |
|---|---|---|---|---|---|
| 1.1 | 5.30 | 11 | 4.41 | 11 | 4.12 |
| 1.2 | 9.13 | 5 | 10.72 | 2 | 10.72 |
| 2.1 | 10.00 | 3 | 10.68 | 4 | 10.68 |
| 2.2 | 2.29 | 13 | 2.30 | 13 | 2.28 |
| 3.1 | 9.12 | 6 | 7.97 | 7 | 9.33 |
| 3.2 | 6.03 | 9 | 5.13 | 10 | 5.13 |
| 4.1 | 8.94 | 7 | 9.36 | 6 | 10.02 |
| 4.2 | 8.21 | 8 | 7.72 | 8 | 7.72 |
| 5.1 | 9.50 | 4 | 9.79 | 5 | 10.38 |
| 5.2 | 5.99 | 10 | 5.42 | 9 | 5.42 |
| 6.1 | 3.46 | 12 | 3.21 | 12 | 2.56 |
| 6.2 | 10.31 | 1 | 10.72 | 2 | 10.72 |
| 7.1 | 10.01 | 2 | 10.89 | 1 | 9.24 |
| 7.2 | 1.72 | 14 | 1.67 | 14 | 1.69 |

By comparing the results of these two assignment manners, we can draw the following conclusions.

Firstly, by comparing the CREE data in Table 3 and Tables 5 and 6, it can be seen that the cross-efficiencies of the 14 sub-DMUs are the same as the cross-efficiencies of the 7 sub-DMUs in stages 1 and 2. This demonstrates that, upon transforming of the seven DMUs with two-stage series structure into 14 sub-DMUs, the relationship between the cross-efficiencies of the stages remains unchanged. This further confirms the correctness of the conclusion drawn in Theorem 1.

Secondly, we examine the sum of the allocations of stages 1 and 2 with these two allocation manners. We find that, based on the three different cooperative game solution concepts, the sum of the revenue allocated to stages 1 and 2 is always invariable regardless of the selected allocation manners, i.e., with these two distribution manners, the revenue allocated to stages 1 and 2 is always $R_1 = 56.32$ and $R_2 = 43.68$, respectively. This result is in accordance with Theorem 4. Therefore, we verify the correctness of Theorem 4 from an illustration perspective.

Finally, comparing Tables 4 and 7, it is easy to observe that under the two allocation schemes, the assignment results according to the nucleolus among the 14 sub-DMUs remain unchanged, whereas the distribution results based on the least core and the Shapley value do not exhibit such characteristics. This indicates that the nucleolus is more stable than the least core and the Shapley value in the revenue allocation problem.

**4.2 An empirical application**

In this subsection, we utilize our proposed revenue allocation method to an empirical problem, i.e., we distribute the revenue among 17 bank branches of China Construction Bank in Anhui Province. We first provide a brief background of the problem and present the associated data.

**4.2.1 The background and data**

China Construction Bank is one of the country's largest state-owned banks. We select 17 bank branches of China Construction Bank in Anhui Province, and adopt the production data first applied by Yang et al. (2011), where they studied the efficiency evaluation and production possible sets for the 17 bank branches using a chain structure DEA model. Each bank branch acts as a two-stage production organization that consumes inputs such as fixed assets ($X_1$), employees ($X_2$), and expenses ($X_3$) to produce intermediate products such as credit ($Z_1$) and interbank loans ($Z_2$) in the first stage. In the second stage, these products are consumed to produce the final outputs of the whole process, such as individuals, organizations or large firms profit from loans and investments, denoted as $Y_1$ and $Y_2$, respectively. The specific raw data and the inputs, intermediate products, and final outputs variables of the bank branch are listed in Table 8.

In the literature, most studies on revenue allocation problems are based on the principle of maximum or constant efficiency to establish a model to maximize the benefits of the collective production. In this empirical problem, the bank's head office needs to give each branch the corresponding rewards according to their performance to motivate them to work harder in their subsequent production activities. Different from other studies, this paper aims to propose a fair and reasonable reward allocation based on the previous contributions of each branch, in order to inspire them to perform better in the future. Thus, our method evaluates participants' performance by using DEA and cooperative games. To ensure fairness, we choose CREE evaluation with peer evaluation as our efficiency score, combined with the competition and cooperation of the participants in the production process to determine the final allocation of the revenue. Therefore, our model is suitable for this empirical problem. We consider each branch as a two-stage series structure that encompasses the production process and the profitability process, assuming that the revenue to be allocated is R = 1000 units.

**Table 8** Raw data of the 17 bank branches.

| Evaluator $DMU_j$ | Branches | Input | | | Intermediate products | | Output | |
|---|---|---|---|---|---|---|---|---|
| | | $X_1(10^8)$ | $X_2(10^3)$ | $X_3(10^8)$ | $Z_1(10^8)$ | $Z_2(10^8)$ | $Y_1(10^8)$ | $Y_2(10^8)$ |
| 1 | Hefei | 1.0168 | 1.221 | 1.2215 | 166.9755 | 8.3098 | 122.1954 | 3.7569 |
| 2 | Bengbu | 0.5915 | 0.611 | 0.4758 | 50.1164 | 1.7634 | 19.4829 | 0.6600 |
| 3 | Huainan | 0.7237 | 0.645 | 0.6061 | 48.2831 | 3.4098 | 34.412 | 0.7713 |
| 4 | Huaibei | 0.5150 | 0.486 | 0.3763 | 35.0704 | 2.348 | 15.2804 | 0.3203 |
| 5 | Maanshan | 0.4775 | 0.526 | 0.3848 | 49.9174 | 5.4613 | 34.9897 | 0.8430 |
| 6 | Tongling | 0.6125 | 0.407 | 0.3407 | 23.1052 | 1.2413 | 32.5778 | 0.4616 |
| 7 | Wuhu | 0.7911 | 0.708 | 0.4407 | 39.4590 | 1.1485 | 30.2331 | 0.6732 |
| 8 | Anqing | 1.2363 | 0.713 | 0.5547 | 37.4954 | 4.0825 | 20.6013 | 0.4864 |
| 9 | Huangshan | 0.4460 | 0.443 | 0.3419 | 20.9846 | 0.6897 | 8.6332 | 0.1288 |
| 10 | Fuyang | 1.2481 | 0.638 | 0.4574 | 45.0508 | 1.7237 | 9.2354 | 0.3019 |
| 11 | Suzhou | 0.7050 | 0.575 | 0.4036 | 38.1625 | 2.2492 | 12.0171 | 0.3138 |
| 12 | Chuzhou | 0.6446 | 0.432 | 0.4012 | 30.1676 | 2.3354 | 13.813 | 0.3772 |
| 13 | Luan | 0.7239 | 0.51 | 0.3709 | 26.5391 | 1.3416 | 5.0961 | 0.1453 |
| 14 | Xuancheng | 0.5538 | 0.442 | 0.3555 | 22.2093 | 0.9886 | 13.6085 | 0.3614 |
| 15 | Chizhou | 0.3363 | 0.322 | 0.2334 | 16.1235 | 0.4889 | 5.9803 | 0.0928 |
| 16 | Chaohu | 0.6678 | 0.423 | 0.3471 | 22.1848 | 1.1767 | 9.2348 | 0.2002 |
| 17 | Bozhou | 0.3418 | 0.256 | 0.1594 | 13.4364 | 0.4064 | 2.5326 | 0.0057 |

**4.2.2 Calculation procedure and analysis of the results**

According to the aforementioned method, we first convert the 17 bank branches with two stages into 34 single-stage sub-DMUs. Subsequently, we use the aggressive CREE model to obtain a unique CREE matrix of these 34 sub-DMUs. Additionally, we calculate the CREE scores of the alliances of the first and second stages separately, as expressed in Tables 9 and 10, respectively.

Since the CREE matrix of 34 sub-DMUs is 34×34 dimensions, we are unable to list it here due to space constraints. However, Theorem 1 and the numerical example illustrate that such a large matrix can be divided into two 17×17-dimensional small matrices, as listed in Table 9 and Table 10, without changing the values.

**Table 9** The cross-efficiency matrix of 17 DMUs in stage 1.

| Evaluator $DMU_j$ | Targeted DMU | | | | | | | | | | | | | | | | |
|---|---|---|---|---|---|---|---|---|---|---|---|---|---|---|---|---|---|
| | 1.1 | 2.1 | 3.1 | 4.1 | 5.1 | 6.1 | 7.1 | 8.1 | 9.1 | 10.1 | 11.1 | 12.1 | 13.1 | 14.1 | 15.1 | 16.1 | 17.1 |
| 1.1 | 1 | 0.44 | 0.41 | 0.41 | 0.64 | 0.23 | 0.24 | 0.18 | 0.24 | 0.2 | 0.33 | 0.29 | 0.22 | 0.23 | 0.24 | 0.2 | 0.19 |
| 2.1 | 1 | 0.77 | 0.58 | 0.68 | 0.95 | 0.5 | 0.66 | 0.49 | 0.45 | 0.72 | 0.69 | 0.55 | 0.52 | 0.46 | 0.51 | 0.47 | 0.62 |
| 3.1 | 1 | 0.54 | 0.63 | 0.59 | 1 | 0.43 | 0.35 | 0.55 | 0.3 | 0.47 | 0.52 | 0.62 | 0.38 | 0.35 | 0.31 | 0.39 | 0.33 |
| 4.1 | 1 | 0.76 | 0.59 | 0.69 | 1 | 0.5 | 0.64 | 0.52 | 0.44 | 0.71 | 0.7 | 0.56 | 0.52 | 0.45 | 0.5 | 0.47 | 0.61 |
| 5.1 | 0.48 | 0.26 | 0.4 | 0.4 | 1 | 0.18 | 0.13 | 0.29 | 0.14 | 0.12 | 0.28 | 0.32 | 0.16 | 0.16 | 0.13 | 0.15 | 0.1 |
| 6.1 | 1 | 0.76 | 0.59 | 0.69 | 1 | 0.5 | 0.64 | 0.52 | 0.44 | 0.71 | 0.7 | 0.56 | 0.52 | 0.45 | 0.5 | 0.47 | 0.61 |
| 7.1 | 1 | 0.77 | 0.58 | 0.68 | 0.95 | 0.5 | 0.66 | 0.49 | 0.45 | 0.72 | 0.69 | 0.55 | 0.52 | 0.46 | 0.51 | 0.47 | 0.62 |
| 8.1 | 1 | 0.54 | 0.63 | 0.59 | 1 | 0.43 | 0.35 | 0.55 | 0.3 | 0.47 | 0.52 | 0.62 | 0.38 | 0.35 | 0.31 | 0.39 | 0.33 |
| 9.1 | 1 | 0.77 | 0.58 | 0.68 | 0.95 | 0.5 | 0.66 | 0.49 | 0.45 | 0.72 | 0.69 | 0.55 | 0.52 | 0.46 | 0.51 | 0.47 | 0.62 |
| 10.1 | 1 | 0.77 | 0.58 | 0.68 | 0.95 | 0.5 | 0.66 | 0.49 | 0.45 | 0.72 | 0.69 | 0.55 | 0.52 | 0.46 | 0.51 | 0.47 | 0.62 |
| 11.1 | 1 | 0.76 | 0.59 | 0.69 | 1 | 0.5 | 0.64 | 0.52 | 0.44 | 0.71 | 0.7 | 0.56 | 0.52 | 0.45 | 0.5 | 0.47 | 0.61 |
| 12.1 | 1 | 0.54 | 0.63 | 0.59 | 1 | 0.43 | 0.35 | 0.55 | 0.3 | 0.47 | 0.52 | 0.62 | 0.38 | 0.35 | 0.31 | 0.39 | 0.33 |
| 13.1 | 1 | 0.76 | 0.59 | 0.69 | 1 | 0.5 | 0.64 | 0.52 | 0.44 | 0.71 | 0.7 | 0.56 | 0.52 | 0.45 | 0.5 | 0.47 | 0.61 |
| 14.1 | 1 | 0.77 | 0.58 | 0.68 | 0.95 | 0.5 | 0.66 | 0.49 | 0.45 | 0.72 | 0.69 | 0.55 | 0.52 | 0.46 | 0.51 | 0.47 | 0.62 |
| 15.1 | 1 | 0.77 | 0.58 | 0.68 | 0.95 | 0.5 | 0.66 | 0.49 | 0.45 | 0.72 | 0.69 | 0.55 | 0.52 | 0.46 | 0.51 | 0.47 | 0.62 |
| 16.1 | 1 | 0.76 | 0.59 | 0.69 | 1 | 0.5 | 0.64 | 0.52 | 0.44 | 0.71 | 0.7 | 0.56 | 0.52 | 0.45 | 0.5 | 0.47 | 0.61 |
| 17.1 | 1 | 0.77 | 0.58 | 0.68 | 0.95 | 0.5 | 0.66 | 0.49 | 0.45 | 0.72 | 0.69 | 0.55 | 0.52 | 0.46 | 0.51 | 0.47 | 0.62 |
| Average CREE | 0.97 | 0.68 | 0.57 | 0.64 | 0.96 | 0.45 | 0.54 | 0.48 | 0.39 | 0.61 | 0.62 | 0.54 | 0.46 | 0.41 | 0.43 | 0.42 | 0.51 |

**Table 10** The cross-efficiency matrix of 17 DMUs in stage 2.

| Evaluator $DMU_j$ | Targeted DMU | | | | | | | | | | | | | | | | |
|---|---|---|---|---|---|---|---|---|---|---|---|---|---|---|---|---|---|
| | 1.2 | 2.2 | 3.2 | 4.2 | 5.2 | 6.2 | 7.2 | 8.2 | 9.2 | 10.2 | 11.2 | 12.2 | 13.2 | 14.2 | 15.2 | 16.2 | 17.2 |
| 1.2 | 1 | 0.58 | 0.57 | 0.34 | 0.44 | 0.85 | 0.76 | 0.34 | 0.27 | 0.3 | 0.33 | 0.42 | 0.24 | 0.72 | 0.26 | 0.39 | 0.02 |
| 2.2 | 1 | 0.71 | 0.57 | 0.34 | 0.44 | 0.85 | 1 | 0.34 | 0.34 | 0.34 | 0.33 | 0.42 | 0.24 | 0.77 | 0.33 | 0.39 | 0.02 |
| 3.2 | 1 | 0.58 | 0.74 | 0.43 | 0.77 | 1 | 0.79 | 0.6 | 0.3 | 0.3 | 0.37 | 0.56 | 0.25 | 0.74 | 0.28 | 0.42 | 0.04 |
| 4.2 | 1 | 0.58 | 0.74 | 0.43 | 0.77 | 1 | 0.79 | 0.6 | 0.3 | 0.3 | 0.37 | 0.56 | 0.25 | 0.74 | 0.28 | 0.42 | 0.04 |
| 5.2 | 1 | 0.58 | 0.74 | 0.43 | 0.77 | 1 | 0.79 | 0.6 | 0.3 | 0.3 | 0.37 | 0.56 | 0.25 | 0.74 | 0.28 | 0.42 | 0.04 |
| 6.2 | 0.52 | 0.28 | 0.38 | 0.25 | 0.24 | 1 | 0.54 | 0.19 | 0.29 | 0.15 | 0.2 | 0.23 | 0.14 | 0.43 | 0.26 | 0.3 | 0.04 |
| 7.2 | 0.56 | 0.42 | 0.38 | 0.23 | 0.24 | 0.63 | 1 | 0.19 | 0.32 | 0.2 | 0.2 | 0.22 | 0.14 | 0.52 | 0.32 | 0.29 | 0.02 |
| 8.2 | 1 | 0.58 | 0.74 | 0.43 | 0.77 | 1 | 0.79 | 0.6 | 0.3 | 0.3 | 0.37 | 0.56 | 0.25 | 0.74 | 0.28 | 0.42 | 0.04 |
| 9.2 | 0.56 | 0.42 | 0.38 | 0.25 | 0.24 | 1 | 1 | 0.19 | 0.48 | 0.2 | 0.2 | 0.23 | 0.14 | 0.52 | 0.46 | 0.3 | 0.24 |
| 10.2 | 1 | 0.71 | 0.57 | 0.34 | 0.44 | 0.85 | 1 | 0.34 | 0.34 | 0.34 | 0.33 | 0.42 | 0.24 | 0.77 | 0.33 | 0.39 | 0.02 |
| 11.2 | 1 | 0.58 | 0.74 | 0.43 | 0.77 | 1 | 0.79 | 0.6 | 0.3 | 0.3 | 0.37 | 0.56 | 0.25 | 0.74 | 0.28 | 0.42 | 0.04 |
| 12.2 | 1 | 0.58 | 0.74 | 0.43 | 0.77 | 1 | 0.79 | 0.6 | 0.3 | 0.3 | 0.37 | 0.56 | 0.25 | 0.74 | 0.28 | 0.42 | 0.04 |
| 13.2 | 1 | 0.58 | 0.74 | 0.43 | 0.77 | 1 | 0.79 | 0.6 | 0.3 | 0.3 | 0.37 | 0.56 | 0.25 | 0.74 | 0.28 | 0.42 | 0.04 |
| 14.2 | 1 | 0.67 | 0.62 | 0.37 | 0.49 | 1 | 1 | 0.38 | 0.38 | 0.33 | 0.34 | 0.45 | 0.24 | 0.78 | 0.36 | 0.41 | 0.07 |
| 15.2 | 0.56 | 0.42 | 0.38 | 0.25 | 0.24 | 1 | 1 | 0.19 | 0.48 | 0.2 | 0.2 | 0.23 | 0.14 | 0.52 | 0.46 | 0.3 | 0.24 |
| 16.2 | 1 | 0.58 | 0.74 | 0.43 | 0.77 | 1 | 0.79 | 0.6 | 0.3 | 0.3 | 0.37 | 0.56 | 0.25 | 0.74 | 0.28 | 0.42 | 0.04 |
| 17.2 | 0.56 | 0.42 | 0.38 | 0.25 | 0.24 | 1 | 1 | 0.19 | 0.48 | 0.2 | 0.2 | 0.23 | 0.14 | 0.52 | 0.46 | 0.3 | 0.24 |
| Average CREE | 0.87 | 0.54 | 0.6 | 0.35 | 0.54 | 0.95 | 0.86 | 0.42 | 0.34 | 0.27 | 0.31 | 0.43 | 0.21 | 0.67 | 0.32 | 0.38 | 0.07 |

Following the algorithm proposed by us, for simplicity, we adopt the second allocation method. Firstly, based on the formula (6) and Theorem 1, we calculate the payoff values of the $2^{17}-1$ sub-coalitions through the CREE matrix and allocate the revenue R to the two stages $N_1$ and $N_2$, obtaining the results $R_1 = 517$ and $R_2 = 483$. From Theorem 4, we know that the nucleolus allocation results for the two distribution modes are stable. Therefore, we use the nucleolus to redistribute the revenues $R_1$ and $R_2$ among the alliances $N_1$ and $N_2$, respectively. The specific allocation results are represented in column 4 of Tables 11 and 12.

Each sub-DMU has 17 CREE values in the CREE matrix, for which we use the arithmetic averaging method to aggregate the CREE values and obtain the average CREE values. These values are then used to rank each sub-DMU, which is then compared with the ranking of the assigned revenue value obtained by the model.

**Table 11** Average CREE, comparison value, allocation result and rank of stage 1.

| DMUs | Stage 1 | | | | |
|---|---|---|---|---|---|
| | Average CREE | Comparison value | Allocation result | Rank(allocation) | Rank(efficiency) |
| 1.1 | 0.97 | 47.63 | 47.63 | 1 | 1 |
| 2.1 | 0.68 | 33.39 | 36.7 | 3 | 3 |
| 3.1 | 0.57 | 27.99 | 30.09 | 8 | 7 |
| 4.1 | 0.63 | 30.93 | 32.98 | 6 | 4 |
| 5.1 | 0.96 | 47.14 | 47.63 | 1 | 2 |
| 6.1 | 0.45 | 22.1 | 23.72 | 14 | 13 |
| 7.1 | 0.54 | 26.52 | 31.2 | 7 | 8 |
| 9.1 | 0.39 | 19.15 | 21.39 | 17 | 17 |
| 10.1 | 0.61 | 29.95 | 34.32 | 4 | 6 |
| 11.1 | 0.62 | 30.44 | 33.22 | 5 | 5 |
| 12.1 | 0.54 | 26.52 | 29.3 | 10 | 8 |
| 13.1 | 0.46 | 22.59 | 24.95 | 12 | 12 |
| 14.1 | 0.41 | 20.13 | 21.77 | 16 | 16 |
| 15.1 | 0.43 | 21.11 | 24.07 | 13 | 14 |
| 16.1 | 0.42 | 20.62 | 22.34 | 15 | 15 |
| 17.1 | 0.51 | 25.04 | 29.37 | 9 | 10 |

**Table 12** Average CREE, comparison value, allocation result and rank of stage 2.

| DMUs | Stage 2 | | | | |
|---|---|---|---|---|---|
| | Average CREE | Comparison value | Allocation result | Allocation rank | Average CREE rank |
| 1.2 | 0.87 | 42.72 | 47.67 | 1 | 2 |
| 2.2 | 0.55 | 27.01 | 33.49 | 7 | 6 |
| 3.2 | 0.6 | 29.46 | 35.21 | 6 | 5 |
| 4.2 | 0.36 | 17.68 | 20.33 | 12 | 11 |
| 5.2 | 0.54 | 26.52 | 36.85 | 4 | 7 |
| 6.2 | 0.95 | 46.65 | 47.67 | 1 | 1 |
| 7.2 | 0.86 | 42.23 | 47.67 | 1 | 3 |
| 8.2 | 0.42 | 20.62 | 28.38 | 8 | 9 |
| 9.2 | 0.34 | 16.7 | 22.69 | 10 | 12 |
| 10.2 | 0.27 | 13.26 | 17.28 | 15 | 15 |
| 11.2 | 0.31 | 15.22 | 17.76 | 14 | 14 |
| 12.2 | 0.43 | 21.11 | 26.85 | 9 | 8 |
| 13.2 | 0.22 | 10.8 | 11.7 | 16 | 16 |
| 14.2 | 0.67 | 32.9 | 36.02 | 5 | 4 |
| 15.2 | 0.32 | 15.71 | 22.16 | 11 | 13 |
| 16.2 | 0.38 | 18.66 | 19.98 | 13 | 10 |
| 17.2 | 0.07 | 3.44 | 11.29 | 17 | 17 |

The reward allocations shown in column 4 of Tables 11 and 12 can be compared with the corresponding average CREE values for 34 sub-DMUs in the second column of Tables 11 and 12, as well as the ranking given in the sixth column of both tables. Upon comparing the ranking of allocation results with the corresponding ranking of average CREE, it is evident that the two results are generally consistent, suggesting that our method of assigning rewards is proportional to the size of CREE. In other words, the corresponding reward distribution based on the peer evaluation is accurately reflected.

In order to illustrate the relationship between the average cross-efficiencies and the allocation results in a more accurate and intuitive way, we present them using line charts. Since the average CREE value is small and the assigned value is relatively large, we uniformly expand the average CREE values by multiplying them by the same coefficient for comparison within the same range. For example, the maximum average CREE value of the 34 sub-DMUs is 0.9694, and its corresponding revenue allocation value is 47.6322. To ensure the comparison and assignment of $DMU_{1.1}$ are equal, we take the coefficient as 47.6322/0.9694 = 49.1357. The comparison values of the other 33 sub-DMUs can be found in the third column of Tables 5 and 6. By comparing the assigned and the comparison values, we can conclude that the changing trends of the comparative values and the allocation values of the 34 sub-DMUs are completely consistent, as shown in Figure 4. This finding confirms that our allocation results are in line with the cross-efficiencies.

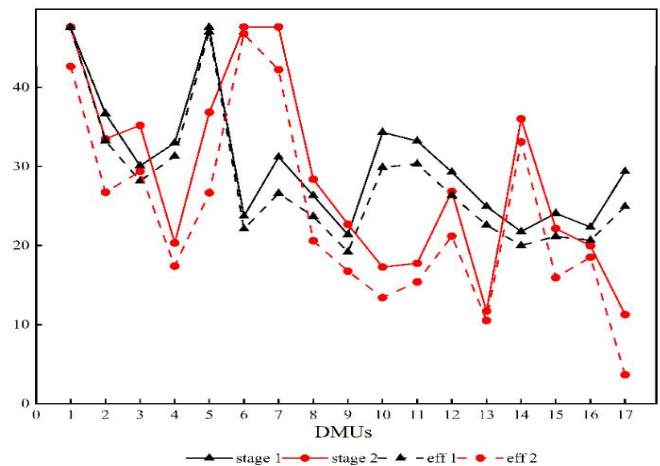

**Figure 4** Resource allocation scheme

Next, we conduct a more detailed analysis based on the figures of revenue distribution. Tables 11 and 12 show that $DMU_1$, the Hefei branch, has the highest distribution in both the first and second stages, which indicates that the Hefei branch, as a provincial branch, is the top performer among the 17 branches in both the production and the profitability processes.

In the first stage, $DMU_5$, the Maanshan branch, performs exceptionally well, tying for the first place with the Hefei branch in terms of distribution result, both as 47.63. In the second stage, $DMU_{5.2}$ is the seventh and, after a comprehensive analysis, it scores the fourth in the profitability process, earning a distribution amount of 36.85.

On the other hand, $DMU_{9.1}$ has the lowest average CREE and allocation in the first stage, indicating a weak performance in the credit and the interbank loaning. However, in the second stage, $DMU_{9.2}$ improves and ranks the twelfth among the 17 sub-DMUs. Thus, it is clear that $DMU_9$ should prioritize the deposit absorption and the interbank loaning in future work.

The $DMU_{6.1}$ has a lower ranking of the fourteenth in the first stage, but its sub-DMU $DMU_{6.2}$ ranks the first in the second stage and obtains the same distribution amount of 47.67 as $DMU_{1.2}$ and $DMU_{7.2}$. This reveals that $DMU_6$ performs poorly in the credit and the interbank loans, yet remarkably well in terms of profitability in the second stage, particularly in the use of deposits and loans. This is an admirable achievement and its experience is worth learning from.

On the contrary, $DMU_{17.1}$ is placed the ninth in the first stage among the 17 sub-DMUs, but its second stage $DMU_{17.2}$ receives the lowest distribution. This shows that $DMU_{17}$ performs poorly in terms of profitability, and it should learn lessons from both $DMU_6$ and $DMU_9$.

## 5. Conclusions

The issue of devising a fair and reasonable revenue allocation scheme among entities with multiple stages is a common challenge faced by managers in the real world. When entities cooperate with each other to form an alliance, it is essential to ensure both the individual rationality and the overall fairness in the revenue allocation scheme. Such a scheme not only ensures the stability and the sustainability of the alliance but also is accepted by all its participants.

To effectively address this situation and achieve the desired allocation schemes, we propose a two-stage series structure to enable participants to cooperate and form an alliance. This alliance will be managed by a centralized manager who will decide the common revenue to be allocated among the participants. Since the allocation results will not be used in the later production process, the allocation scheme will depend on the participants' prior performances. To this end, we have developed allocation schemes combining CREE with cooperative concepts such as the nucleolus, the least core, and the Shapley value. The concrete common revenue allocation procedure is as follows.

By extending the vectors of the initial input, the intermediate product as well as the final output, we can convert $n$ DMUs with two-stage structure into $2n$ single-stage sub-DMUs, and then derive the CREE matrix of $2n$ sub-stages. Based on CREE, we design a characteristic function and analyze its properties, such as the super-additivity and the core non-emptiness. Furthermore, we propose two ex post allocation manners, namely the direct allocation mode and the secondary allocation mode, and demonstrate that the sum of the revenue allocated to all DMUs remains unchanged at each stage, regardless of the chosen allocation manner and the cooperative solution concept. Finally, we illustrate the calculation procedure and usefulness of our proposed approach via a numerical example and an empirical application..

We summarize the following advantages of our method compared with previous ones in the literature: (1) Our method takes into account the internal structures of the production organization and simultaneously calculates the revenue distribution between and within the production organization. (2) By extending the vectors of the initial input, the intermediate product and the final output, we present a CREE evaluation method for a two-stage series production organization structure; this offers a new approach to dealing with CREE problems of multi-stage and complex structures. (3) We use CREE and the cooperative game solution concepts of the nucleolus, the least core and the Shapley value to allocate the revenue, thus not only reflecting the overall fairness of the centralized decision-making problem, but also considering the individual rationality of the participants from the perspectives of marginal contribution and satisfaction. Therefore, the allocation scheme derived from our method is comprehensive, flexible and can be accepted by all participants.

Going forward, two potential research directions can be derived from this study. Firstly, the two-stage structure of DMUs can be extended to the more complex network structures to better reflect real-life scenarios. Secondly, for the commonly encountered issue of revenue distribution, further research should not only take into account the marginal contribution and contentment of the alliance members, but also contemplate the implementation of penalty measures.

## Acknowledgement

This work was supported by National Natural Science Foundation of China (Grant No. 72001207).